\documentclass[]{elsarticle}
\usepackage{amsmath}
\usepackage{color}
\usepackage{ulem}
\usepackage{textcomp}
\usepackage{upgreek}

\newcommand{\xalpha}{\upalpha}

\title{Iron single crystal growth from a lithium-rich melt}

\author[ep6]{M. Fix}
\ead{manuel.fix@physik.uni-augsburg.de}

\author[freiberg]{H. Schumann}
\ead{Helge.Schumann@iww.tu-freiberg.de}

\author[ssc]{S. G. Jantz}
\ead{stephan.jantz@physik.uni-augsburg.de}

\author[ep6]{F. Breitner}
\ead{franziska.breitner@physik.uni-augsburg.de}

\author[freiberg]{A. Leineweber}
\ead{Andreas.Leineweber@iww.tu-freiberg.de}

\author[ep6]{A.\,Jesche\corref{cor1}}
\ead{anton.jesche@physik.uni-augsburg.de}

\cortext[cor1]{anton.jesche@physik.uni-augsburg.de}

\address[ep6]{EP VI, Center for Electronic Correlations and Magnetism, Institute of Physics, University of Augsburg, D-86159 Augsburg, Germany}

\address[freiberg]{Institute of Materials Science, TU Bergakademie Freiberg, Gustav-Zeuner-Str. 5, D-09599 Freiberg, Germany}

\address[ssc]{Lehrstuhl f\"ur Festk\"orperchemie, Institut f\"ur Physik, Universit\"at Augsburg, D-86159 Augsburg, Germany}

\begin{document}

\begin{abstract}
$\xalpha$-Fe single crystals of rhombic dodecahedral habit were grown from a melt of Li$_{84}$N$_{12}$Fe$_{\sim 3}$.
Crystals of several millimeter along a side form at temperatures around $T \approx  800^\circ$C.
Upon further cooling the growth competes with the formation of Fe-doped Li$_3$N.
The b.c.c. structure and good sample quality of $\xalpha$-Fe single crystals were confirmed by X-ray and electron diffraction as well as magnetization measurements and chemical analysis.
A nitrogen concentration of 90\,ppm was detected by means of carrier gas hot extraction.
Scanning electron microscopy did not reveal any sign of iron nitride precipitates.
\end{abstract}


\maketitle
\section{Introduction}
Iron is one of the most abundant materials in the earth's crust.
As the main ingredient of steel it is still 
- and probably that won't change soon - 
of vital importance as a construction material.
Even after many centuries of application and research elemental Fe is not as well understood as one might think. 
The lattice dynamics in $\xalpha$-Fe, for example, are significantly affected by many-body effects and have been properly modeled only quite recently\,\cite{Leonov2014}.  

The occurrence of structural transitions from $\updelta$-Fe to $\upgamma$-Fe at $T = 1394^\circ$C and $\upgamma$-Fe to $\xalpha$-Fe at $T = 912^\circ$C upon cooling does not allow for the growth of mono-domain $\xalpha$-Fe single crystals from the liquid.
The strain-anneal method\,\cite{Carpenter1921, Kadeckova1967, Lubitz1979, Behr2007} works around this problem and is the standard process for the production of commercially available bulk single crystals of $\xalpha$-Fe.
Comparatively large single crystals can be also grown in form of whiskers\,\cite{Gardner1978}.
Various single crystalline $\xalpha$-Fe nano structures were created, among others: 
thin films by molecular beam epitaxy\,\cite{Prinz1986},
nanowires by pulsed laser deposition\,\cite{Ardabili2005}, 
nanosized crystals by electro deposition in solution\,\cite{Lu1997} or induced by electron radiation\,\cite{Zhang2006}. 

A rhombic dodecahedral habit, as one of the natural forms of a cubic material, was achieved by using a gas-evaporation technique\,\cite{Uyeda1974,Hayashi1977}.
The sample diameter, however, was limited to 200\,nm.  
Here we report on the growth of rhombic dodecahedral $\xalpha$-Fe single crystals with diameters in the range of millimeters.

\section{Methods}
Standard box and tube furnaces were used for vertical and horizontal crucible orientation, respectively.  
Nb and Ta crucibles were machined from commercially available tubes and sheets\,\cite{Canfield2001, Jesche2014c}.  
X-ray diffraction patterns in reflection mode were measured using a Rigaku Miniflex 600 diffractometer (Cu-K$\alpha_{1,2}$, Ni-filter).
Laue-back-reflection patterns were recorded with a digital Dual FDI NTX camera manufactured by Photonic Science (W anode, $U = 15$\,kV, $I = 30$\,mA, beam diameter roughly 1\,mm).
Mosaicity was estimated in transmission mode using a Bruker D8 Venture single crystal diffractometer equipped with a SMART APEXII 4k CCD detector (Mo-K$\alpha$ radiation).
Magnetization measurements were performed using a Quantum Design Magnetic Property Measurement System (MPMS 3). 
The composition of the samples was analyzed by inductively coupled
plasma optical emission spectroscopy (ICP-OES) using a Varian Vista-MPX.
A Scanning Electron Microscope (SEM, Zeiss LEO 1530 Gemini) equipped with an Electron Backscatter Diffraction (EBSD) system (HKL Technology, working distance = 15\,mm, $U = 20$\,kV, step size = 0.8\,\textmu m) was used to determine the crystal orientation.
The nitrogen concentration was measured by carrier gas hot extraction using a Bruker G8 Galileo analyzer.

\section{Single Crystal Growth of $\xalpha$-Fe}
After first $\xalpha$-Fe single crystals were obtained by the \textit{standard} method described in the following (Sec\,\ref{boxsec}), we performed further growth attempts in a horizontal tube furnace in order to study the influence of a temperature gradient (Sec\,\ref{tube}). 

\subsection{Low-gradient box furnace}\label{boxsec}
\begin{figure}
\center
\includegraphics[width=0.99\textwidth]{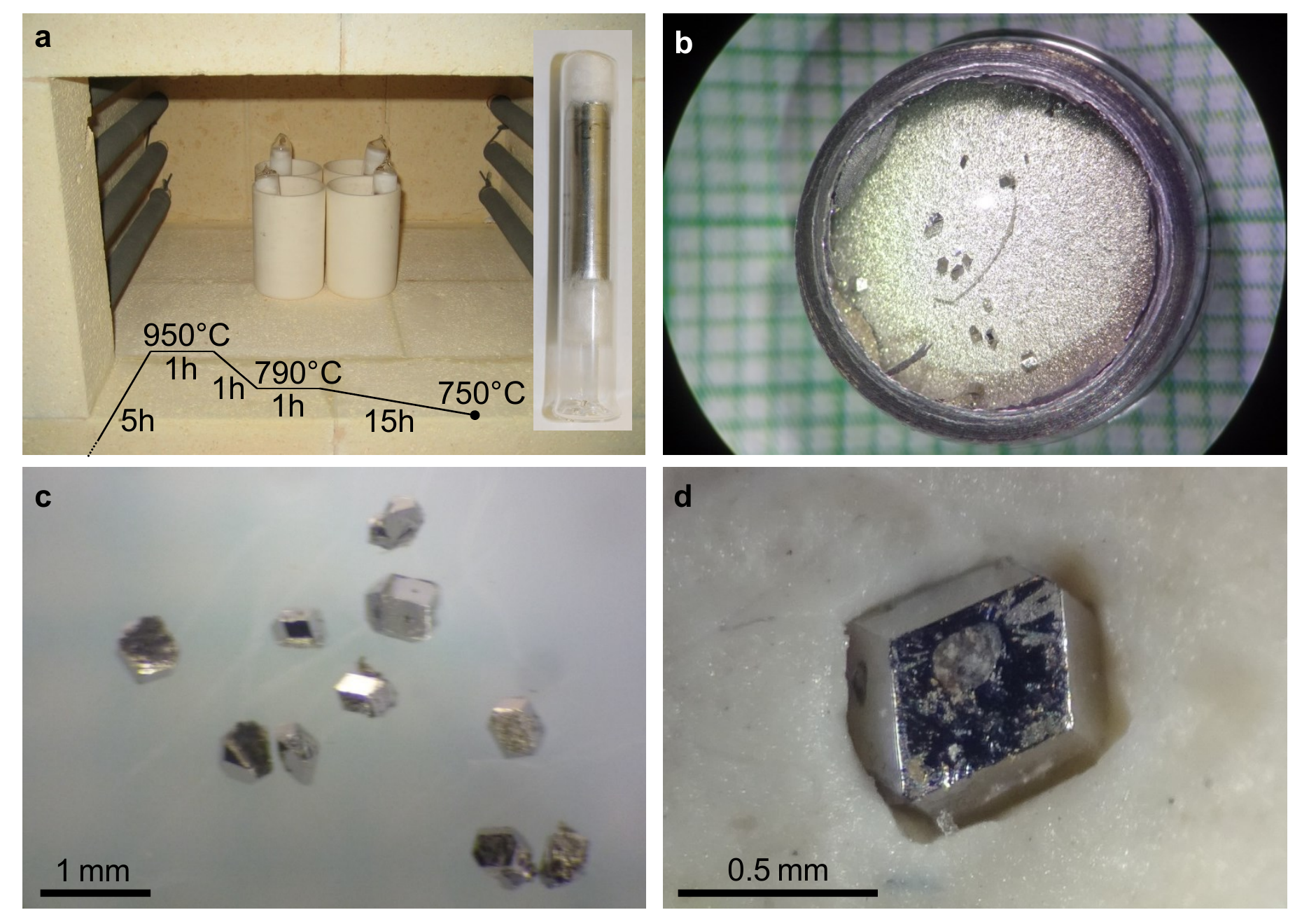}
\caption{Single crystal growth of $\xalpha$-Fe in Nb 1/2-inch crucibles. (a) Box furnace containing four Nb containers held vertically by larger Al$_2$O$_3$ crucibles. The Nb crucibles are sealed in fused silica ampules (inset) to avoid oxidation.
The temperature profile is indicated.
(b) Fe single crystals at the crucible bottom after decanting the flux at $T = 750^\circ$C and opening of the assembly. (c) The obtained crystals show mainly rhombic dodecahedral habit. 
(d) Selected crystal with only minor deviations from a regular rhombic dodecahedron.  
\label{box}}
\end{figure}

Starting materials were Li foil (Alfa Aesar, 99.9\,\%), Fe granules (Alfa Aesar, 99.95\,\%) and Li$_3$N powder (Alfa Aesar, 99.4\,\%).
The materials were mixed in a molar ratio of Li:Li$_3$N:Fe = 48.4:12:2. 
The mixture with a total mass of ${\sim}\,1.5$\,g was placed in a Nb three-cap crucible\,\cite{Canfield2001, Jesche2014c} that was sealed under roughly 0.6\,bar Ar by arc welding.  
The closed Nb crucible was sealed in a fused silica ampule under ${\sim}\,0.2$\,bar Ar to prevent oxidation of Nb (Fig.\,\ref{box}a).
Using a box furnace the mixture was heated to $950^\circ$C over 5\,h, held for 1\,h, cooled to $790^\circ$C over 1\,h and held for 1\,h.
The gradient in this temperature range amounts to roughly 2\,K/cm.
Finally the mixture was cooled to $750^\circ$C over 15\,h. 
At that temperature the liquid was decanted: the hot ampule was transfered into a centrifuge within ${\sim}\,3$\,s and, within ${\sim}\,4$\,s, accelerated to ${\sim}\,1000$\,rpm (corresponding to ${\sim}\,300$\,g). 
After cooling to room temperature (RT) within roughly 15\,mins the Nb ampule was opened in an Ar filled glove box. 
Facetted single crystals were found at the bottom of the Nb crucible (Fig.\,\ref{box}b-d).

\subsection{Horizontal tube furnace}\label{tube}
\begin{figure}[t]
\center
\includegraphics[width=0.99\textwidth]{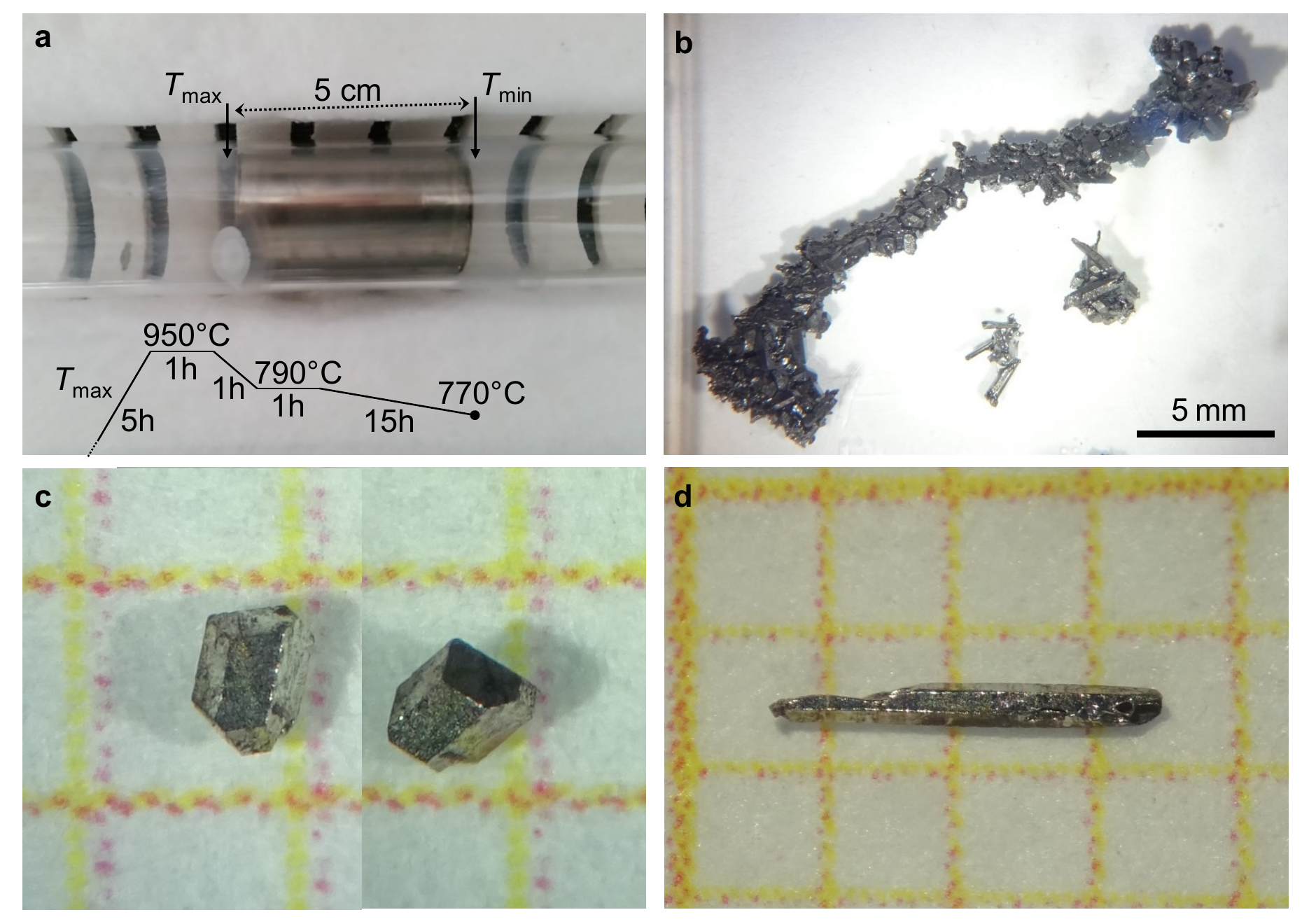}
\caption{Single crystal growth of $\xalpha$-Fe in a horizontal tube furnace. 
(a) Quartz tube holding a Ta crucible. $T_{\rm max}$ marks the center of the heating zone, whereas $T_{\rm min} \approx T_{\rm max} - 25^\circ$C denotes the colder end of the crucible. The temperature profile is indicated.
(b) Conglomerates of single crystals obtained after dissolving the flux in deionized water.
(c) Most of the crystals are of rhombic dodecahedral habit enclosed by \{1\,1\,0\} (left crystal, on mm grid), others do show also $90^\circ$ and $135^\circ$ angles indicating the presence of \{1\,0\,0\} facets (right crystal).
(d) The presence of rod shaped crystals (on mm grid) suggests an increasing tendency to whisker formation in larger temperature gradients. 
\label{tubefig}}
\end{figure}

In addition to the variable temperature gradient of the horizontal furnace, the employed setup allows to work with larger amounts of material: the crucible diameter was doubled to 2.5\,cm. We used Ta instead of Nb as crucible material because it is less sensitive to oxidation (the sealing of the tube furnace is not as good as the one of fused silica ampules). A Ta cap was arc welded to the Ta tube (5\,cm in length) and acts as crucible bottom. 
Starting materials were Li granules (Alfa Aesar, 99\,\%), Fe granules (Alfa Aesar, 99.95\,\%) and Li$_3$N powder (Alfa Aesar, 99.4\,\%).  
The materials were mixed in a molar ratio of Li:Li$_3$N:Fe = 48.4:12:3.6 with a total mass of $10$\,g. 
The mixture was sealed by arc welding a second cap to the top of the Ta tube under roughly 0.6\,bar Ar (Fig.\,\ref{tubefig}a).

One end of the crucible was placed in the center of the heating zone of a horizontal tube furnace. 
The (center of the) furnace was heated to $950^\circ$C over 5\,h, held for 1\,h, cooled to $790^\circ$C over 1\,h, held for 1\,h and finally cooled to $770^\circ$C over 15\,h.
A temperature difference of $\Delta T = 25$\,K was found between the center of the heating zone and the colder end of the Ta crucible.
The corresponding gradient amounts to 5\,K/cm. Note that the gradient within the crucible could be lower.
Before rapid cooling to RT, the lowest temperature in the crucible was $T = 745^\circ$C (at the right hand side), similar to the growth using the box furnace. 
After completing the annealing process and an additional hold for 10\,h the crucible was quickly transfered to the cold part of the quartz tube and cooled down to RT within ${\sim}\,$15\,min.

After cooling, the Ta crucible and the contained product was cut into three parts by using a tube cutter.
Optical inspection of the product did not indicate any obvious inhomogeneities within the crucible. 
The Li-rich flux was subsequently dissolved in water and $\xalpha$-Fe single crystals similar to the ones obtained in the box furnace (see Sec.\,\ref{boxsec}) were found. 
Part of the sample formed larger conglomerates that are loosely connected by flux remnants or sintered together (Fig.\,\ref{tubefig}b). 
The majority of the crystals is of rather isometric, rhombic dodecahedral habit. 
Some of those do also show $90^\circ$ and $135^\circ$ angles that indicate the presence of \{1\,0\,0\} facets (Fig.\,\ref{tubefig}c).
However, a significant number of crystals is rod-shaped as shown in Fig.\,\ref{tubefig}d.
The crossover from isometric to rod or needle shaped is smooth: larger aspect ratios up to ${\sim}\,30$ are found with more or less monotonically decreasing probability.  

Another growth process was performed with the same parameters but slow cooling starting already at $T = 825^\circ$C instead of $T = 790^\circ$C. 
The cooling rate was kept the same and the temperature gradient does not change significantly between those temperatures. 
Before dissolving the flux the product appears similar to the previously obtained one.
The amount of rod-shaped crystals, however, increased significantly and forms the majority of the obtained Fe crystals.
Isometric, rhombic dodecahedra are also present, but in smaller numbers when compared to the previous attempt.

\section{X-ray diffraction}
\begin{figure}
\center
\includegraphics[width=0.99\textwidth]{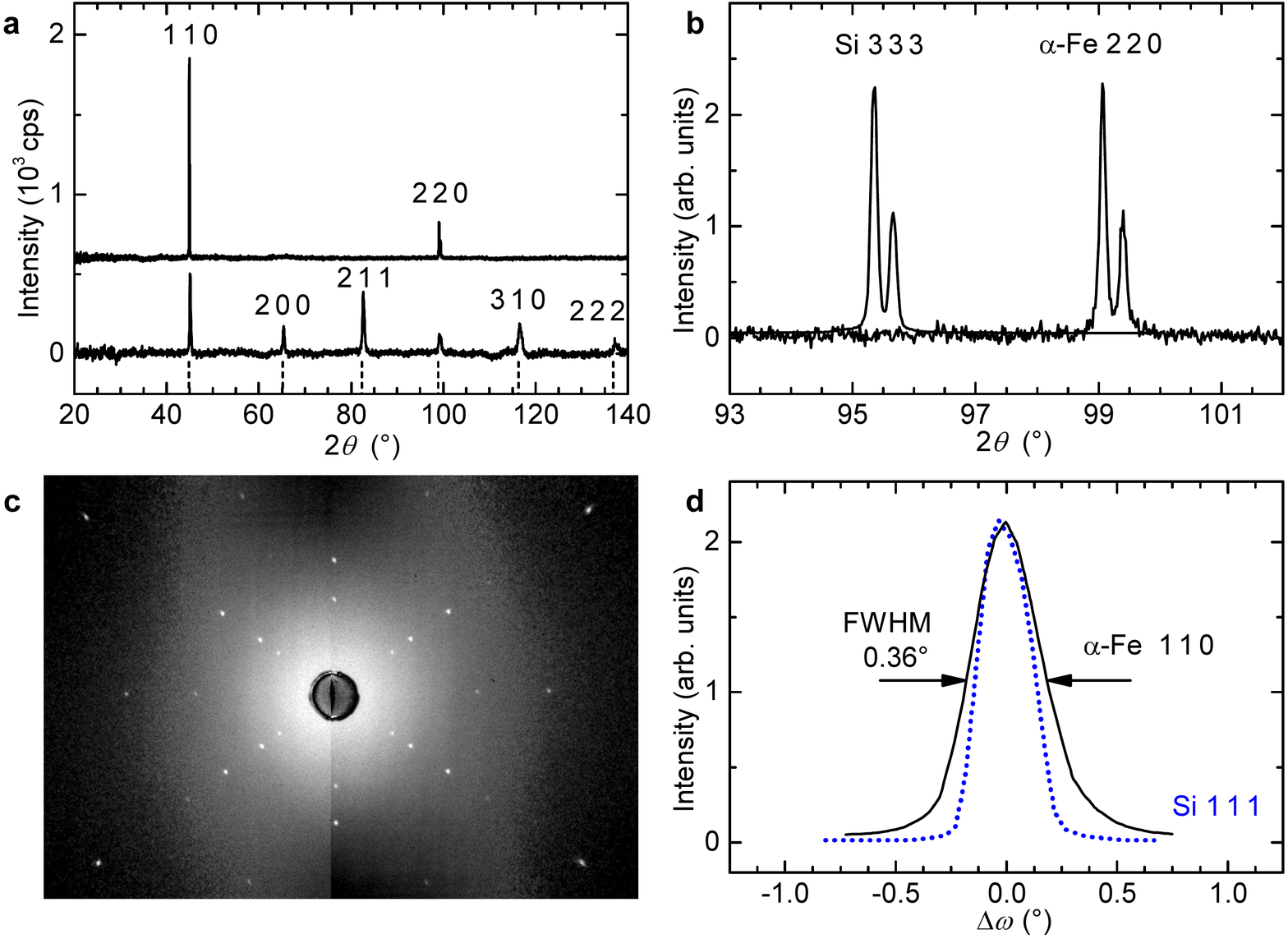}
\caption{X-ray diffraction pattern of $\xalpha$-Fe. (a) One single-crystal in \{$1\,1\,0$\} orientation (upper line) and an assembly of randomly oriented single-crystals (lower line) measured using a powder diffractometer in Bragg-Brentano geometry (Cu-K$\alpha$ radiation). 
Theoretical peak positions (dashed lines) and Miller indices of $\xalpha$-Fe are indicated. 
(b) Enlarged view on $\xalpha$-Fe $2\,2\,0$ and comparison with Si $3\,3\,3$ obtained in identical configuration.
(c) Laue-back-reflection pattern for incident beam parallel to the surface normal of an as-grown facet, that is along [$1\,1\,0$]. All reflections can be indexed based on the b.c.c. lattice of $\xalpha$-Fe.  
(d) 'Rocking curve' for $\xalpha$-Fe $1\,1\,0$ in comparison with Si $1\,1\,1$ (dotted, blue line).
\label{diff}}
\end{figure}

Optical inspection of the obtained crystals (Fig.\,\ref{box}) already revealed significant differences to the anticipated Fe-doped Li$_3$N which forms hexagonal platelets\,\cite{Jesche2014b} and motivated a more careful analysis by means of X-ray diffraction.
The obtained patterns are presented in Fig.\,\ref{diff} and turned out to be consistent with the crystal structure of $\xalpha$-Fe\,\cite{Westgren1922}.
The upper data set in Fig.\,\ref{diff}a was measured by orienting a surface of one single-crystal parallel to the sample holder of a powder diffractometer such that Bragg reflection is possible (that is: surface normal parallel to $\vec k - \vec k_0$)\,\cite{jesche2016}.
Only the $1\,1\,0$ and the $2\,2\,0$ reflections of $\xalpha$-Fe are observed. 
All reflections expected for $\xalpha$-Fe emerge when several single-crystals are randomly oriented on the sample holder (lower curve in Fig.\,\ref{diff}a). 
The obtained lattice parameters of $a = 2.867(3)$\,\r{A} for the oriented single crystal and $a = 2.868(3)$\,\r{A} for the collection of randomly oriented single crystals are in good agreement with literature data (2.867\,\r{A}\,\cite{Westgren1922}, 2.8665(5)\,\r{A}\,\cite{Fuchs1965}).
Fig.\,\ref{diff}b shows an enlarged view of the Fe $2\,2\,0$ reflection. 
The Cu-K$\alpha_{1,2}$ splitting is well resolved and the peak-width comparable to the one obtained for Si $3\,3\,3$ (piece of Si wafer measured in similar configuration). 

A Laue-back-reflection pattern for the incident beam parallel to a surface normal is shown in Fig.\,\ref{diff}c. 
The picture corresponds to a detector area of $13.2 \times 9.4$\,cm$^2$, sample detector distance was 3.9\,cm, exposure time 30\,mins.
A crystal with an exposed surface of roughly 1/4\,mm$^2$ was centered in the 1\,mm diameter beam.
The positions of all measured reflections are in excellent agreement with the pattern calculated for [$1\,1\,0$] orientation parallel to the incident beam. 
The peaks are comparatively sharp and show no indication for enhanced mosaicity or twinning. 
In order to further evaluate the mosaicity, a sample with dimensions of $0.20 \times 0.20 \times 0.18$\,mm$^3$ was selected for single crystal diffraction in transmission mode. 
The measured 'rocking curve' ($\omega$-scan) is shown in Fig.\,\ref{diff}d.
A full width at half maximum (FWHM) of $\Delta \omega_{\rm total} = 0.36^\circ$ was obtained for $\xalpha$-Fe $1\,1\,0$. 
The instrumental broadening is estimated from a corresponding measurement on Si $1\,1\,1$ and amounts to $\Delta \omega_0 = 0.28^\circ$ (dotted, blue line in Fig.\,\ref{diff}d). 
Assuming $\Delta \omega_{\rm sample}^2 + \Delta \omega_{0}^2 =  \Delta \omega_{\rm total}^2$ we find an intrinsic mosaicity of $\Delta \omega_{\rm sample} = 0.23^\circ$ for the $\xalpha$-Fe single crystal.

\section{Chemical analysis}
For ICP-OES analysis several Fe crystals with a total mass of 14\,mg were rinsed in acetone, dried, and dissolved in 4\,ml hydrochloric acid solution (37\,\%).
The mixture was further diluted by adding 46\,ml distilled water.
The concentration of potential impurities was measured against multi-element standards (Alfa Aesar) of 1 and 2\,ppm in solution. 
The combined concentration of all measured impurities, that is elements other than Fe, comprised less than 0.4\,mass\% of the solid samples.
An overview of detected contaminant elements is given in Tab.\,\ref{tab:icp}. Only Cu and Ca were found in larger amounts. 
Whereas Ca is a known impurity in Li and Li source materials\,\cite{Virolainen2016}, the origin of Cu is less obvious. 
However, the measured impurity levels are in accordance with the purity of the starting materials.
Further elements were searched for but were found to be below the resolution limit (of roughly 10\,ppm of the solid Fe sample): Ag, Al, Bi, Cd, Cr, Ga, Na, Pr, Sr.

The N concentration was determined by carrier gas hot extraction using roughly 500\,mg of $\xalpha$-Fe single crystals (synthesized in the horizontal furnace).
A concentration of 90\,ppm (mass ratio) was found. 
The experimentally determined N content in the Fe crystals can provide information about the chemical potential or the N$_2$ partial pressure present during growth of these crystals. 
The known relation between N uptake by $\xalpha$-Fe from N$_2$ gas as a function of the partial pressure\,\cite{Wriedt1987} yields an N$_2$ pressure of 14\,bar.
This value, however, appears very high considering the excess of Li and the highly exothermic reaction between Li and N with a standard enthalpy of formation of $\Delta\,H_f^0\,=\,-165.6\,{\rm kJ(mol\,Li_3N)^{-1}}$ \,\cite{Wang2003}.
An estimate of the equilibrium nitrogen partial pressure in the relevant  thermodynamic parameter range yields an N$_2$ partial pressure close to zero. Therefore, it seems possible that the measured N content in the Fe crystals could, for example, result from surface contaminations.

\begin{table}
\caption{Impurity concentrations in $\xalpha$-Fe (ppm mass ratio). The nitrogen concentration was determined by carrier gas hot extraction. All other values were measured by ICP-OES.} 
\center
\begin{tabular}{c r c}
\hline
\hline
element & conc. & error\\
\hline
B		& 80	&	14	\\
Ba		& 14 	&	4	\\
Ca		& 500	&	13	\\
Co		& 80	&	28	\\
Cu		& 3000	&	90	\\
In		& 200	&	25	\\
Mg		& 20	&	7	\\
Mn		& 140	&	8	\\
Ni		& 90	&	10	\\
Zn		& 62	&	8	\\
\hline
N		& 90	&		\\
\hline
\hline
\end{tabular}
\label{tab:icp}
\end{table}

Furthermore, the magnetization of two single crystals with masses of $m = 0.59$\,mg (box furnace) and $m = 2.44$\,mg (tube furnace), respectively, were measured at $T = 300$\,K in applied fields of up to $\mu_0H = 7$\,T. 
We found a saturation magnetization of 233(12)\,emu/g for the former and 226(5)\,emu/g for the latter.
Both values are slightly enlarged compared to the literature data (217\,emu/g)\,\cite{Crangle1971}.
Enhanced saturation magnetization could indicate the presence of iron nitrides\,\cite{Nakajima1990, Coey1999}.
However, the small concentration of N renders this scenario unlikely, at least when a homogeneous N distribution is considered (Note that roughly 300 single crystals were necessary to perform the carrier gas hot extraction).  
Remnant magnetization and coercivity field were below the resolution limits (4\,emu/g and 40\,Oe, respectively, for the employed setup and sample size).
   
\section{Electron microscopy and diffraction}
\begin{figure}
\center
\includegraphics[width=0.99\textwidth]{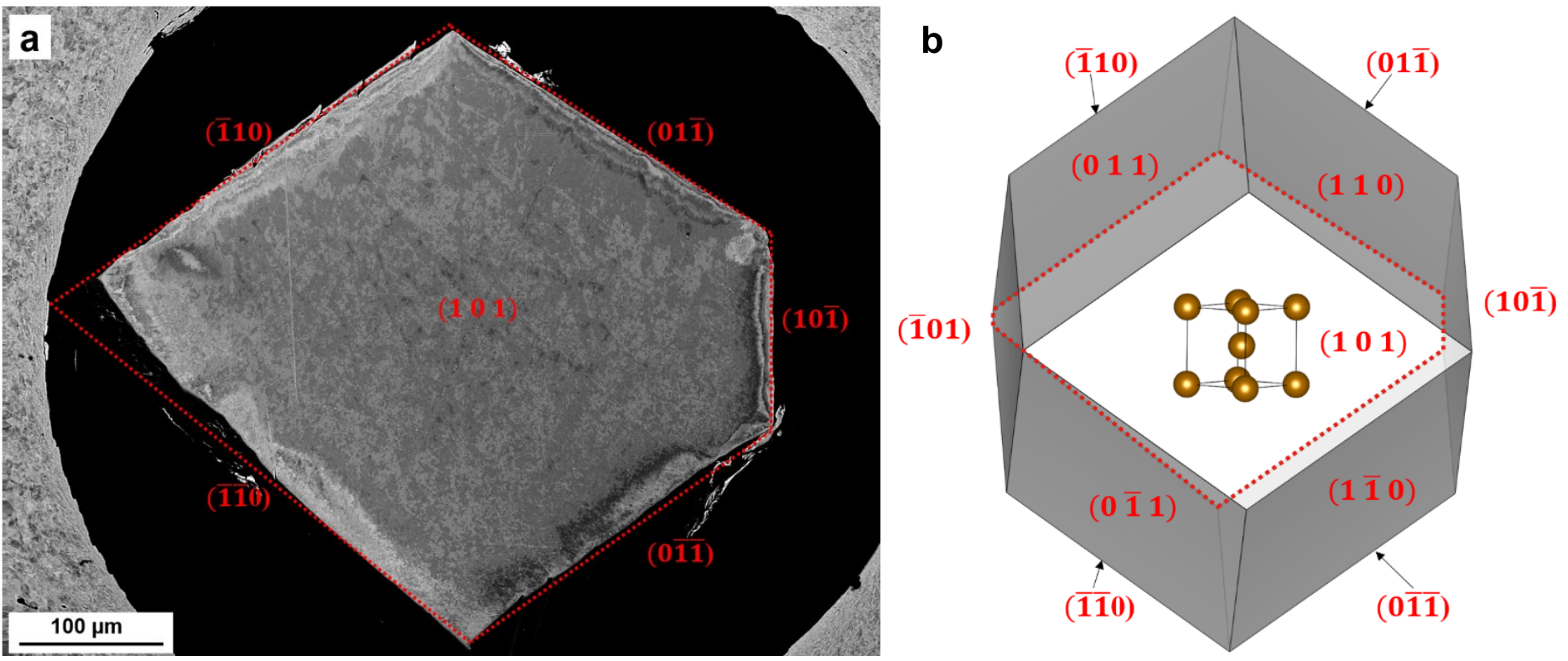}
\caption{Electron backscatter diffraction (EBSD) results on $\xalpha$-Fe single crystals. (a) SEM micrograph of the polished sample selected for EBSD. The red dotted lines show the theoretical directions of the $\{1\,1\,0\}$ lattice plane traces. 
(b) Illustration of the morphology of the iron single crystal with $\{1\,1\,0\}$ cutoff planes (rhombic dodecahedron) and orientation roughly like the measured single crystal. 
The crystal is cut parallel to the sample surface by a near $(1\,0\,1)$ plane. Inset: Orientation of the b.c.c. unit cell.
\label{sem}}
\end{figure}

For SEM analysis, an iron single crystal with a diameter of about {500\,\textmu m} was placed for mechanical support into a hole drilled into an iron plate. 
That arrangement was embedded into PolyFast (Struers GmbH) resin and mechanically ground and polished. 
Fig.\,\ref{sem}a shows a SEM micrograph of this single crystal in backscattered electron contrast after removal of some of its surface by the grinding/polishing. 
The different gray shades caused by local differences in the intensity of the back-scattered electrons do likely originate from preparation effects, especially local deformation at the edges. 

Kikuchi patterns measured from the polished crystal surface are consistent, as expected, with the presence of ${\rm \xalpha}$-Fe in a single, well-defined orientation. 
The visible, approximately rhombic, polished surface of the crystal corresponds to a (0.66\,0.08\,0.75) plane that deviates from (1\,0\,1) by $6^\circ$.
Based on the crystal orientation matrix obtained from the Kikuchi patterns the positions of all $\{1\,1\,0\}$ planes were calculated.
The result is sketched in Fig.\,\ref{sem}b with respect to an ideal rhombic dodecahedron. 
The orientation of the b.c.c. unit cell is shown in the inset. 
The further trace, which bounds ($1\,0\,\bar{1}$) and appears as a vertical line in Fig.\,\ref{sem}a, is also found. 
This truncation is caused by the tilted removal of material from the rhombic dodecahedron. 
Small differences between the observed crystal edges and the calculated plane traces can be explained by additional effects as a shifted aspect ratio formed during the growth of the crystals (see Fig.\,\ref{tubefig}c)\,\cite{Sunagawa2010}
and errors with regard to the EBSD measurements\,\cite{Nolze2007}.

\section{Discussion}
The solubility of Fe in pure Li is well below 1\,\%\,\cite{Massalski1996,Lyublinski1995}. However, it has been shown that small amounts of nitrogen (and also oxygen) do increase the solubility of Fe significantly\,\cite{Lyublinski1995}. 
For example, raising the nitrogen impurity level from 0.05\,wt\% to 0.49\,wt\% leads to an increase in the Fe solubility from 0.17\,wt\,\% to 3.5\,wt\% (at $T = 950^\circ$C, see Table\,2 in\,\cite{Lyublinski1995}).
Extrapolating to a N concentration of 16\,wt\% (present here), we obtain an Fe solubility of ${\sim}\,60$\,wt\%.
Taking into account the large difference in the concentration range, this value is in good agreement with our estimate of $\approx 20$\,wt\% that was determined experimentally by subsequently increasing the amount of Fe in the starting material until the solubility limit is reached.

According to the Fe-N binary phase diagram\,\cite{Massalski1990b}, the presence of N could lead to a primary crystallization of $\upgamma$-Fe that undergoes a eutectoid transformation at $T = 592^\circ$C to produce $\upalpha$-Fe and Fe$_4$N.
However, we would expect the development of several domains at the structural transition ($\upgamma \rightarrow \upalpha$) in contrast to the observed formation of single domain $\upalpha$-Fe. 
Furthermore, no traces of Fe$_4$N were found. 
The affinity of N to Li seems to prevent the formation of $\upgamma$-Fe in the ternary Li-Fe-N melt.

Only about 4\,\% (in the box furnace) of the total amount of Fe forms $\xalpha$-Fe single crystals upon cooling to $T = 750^\circ$C, the vast majority remains dissolved in the Li-rich melt.
A more accurate estimate of the Fe concentration would require significantly more effort since small Fe crystals could pass through the strainer during centrifugation.
Investigating the spin-side of the crucible in more detail is not very helpful in this regard because it is not possible to accurately evaluate the crystallization during rapid cooling (after centrifugation).
In the horizontal tube furnace the fraction of formed $\xalpha$-Fe single crystals amounts to ${\sim}\,$24\,\%, considerably higher when compared to the box furnace attempt. 
This seems reasonable since the samples are not centrifuged and therefore all grown crystals can be retrieved. 
Nevertheless, we cannot rule out the possibility of the larger yield being caused by a higher temperature gradient in the horizontal furnace.

Various growth procedures were performed that aimed at single crystalline Li$_2$(Li$_{1-x}$Fe$_x$)N using methods and parameters similar to the ones described in Sec\,\ref{boxsec}. 
No Fe single crystals were found in those attempts when the mixture was slowly cooled to $T = 710^\circ$C or below.
Instead, a significant amount of Li$_2$(Li$_{0.7}$Fe$_{0.3}$)N forms. 
A rough estimate indicates that the amount of iron in the melt would be fully sufficient for the formation of the latter even when considering the ambiguities in the iron concentration of the melt mentioned above. 
Accordingly, Li$_2$(Li$_{1-x}$Fe$_x$)N does not necessarily grow on the expense of dissolving Fe single crystals.
Rather the Fe solubility in the Li-rich melt could increase upon cooling, i.e. the temperature coefficient of the solubility is negative. 
An instability of Li-Fe-N or Fe-N complexes, which are essential for the solubility of Fe in Li, could become unstable at elevated temperatures.
Further investigations are necessary to gain a better understanding of nucleation, solubility and the microscopic details of the growth process. However, those are beyond the scope of this paper. 
 
To summarize, we have synthesized $\xalpha$-Fe single crystals that grow as millimeter sized rhombic dodecahedra.
As such they grow in their natural crystal habit in accordance with the cubic lattice and in contrast to the formation of Fe whiskers obtained else-wise.
It seems not unlikely to find an optimized flux that allows for larger and cleaner Fe single crystals at even lower temperatures. 

\section*{Acknowledgement}
The authors wish to thank Thilo Kreschel from the Institute of Iron and Steel Technology (TU Bergakademie Freiberg) for performing the carrier gas hot extraction measurements. We thank Theodor Gr\"unwald, Alexander Hartwig, Alexander Herrnberger, Danuta Trojak, and Klaus Wiedenmann for technical assistance. Andrea Mohs is acknowledged for performing the ICP-OES measurements. We thank Wolfgang L\"oser, Wolf Assmus and Peter H\"ohn for useful comments and discussions.
This work was supported by the Deutsche Forschungsgemeinschaft (DFG, German Research Foundation) - Grant No. JE 748/1.

\bibliographystyle{elsarticle-num}

\begin{thebibliography}{10}
\expandafter\ifx\csname url\endcsname\relax
  \def\url#1{\texttt{#1}}\fi
\expandafter\ifx\csname urlprefix\endcsname\relax\def\urlprefix{URL }\fi
\expandafter\ifx\csname href\endcsname\relax
  \def\href#1#2{#2} \def\path#1{#1}\fi

\bibitem{Leonov2014}
I.~Leonov, A.~I. Poteryaev, Y.~N. Gornostyrev, A.~I. Lichtenstein, M.~I.
  Katsnelson, V.~I. Anisimov, D.~Vollhardt, Electronic correlations determine
  the phase stability of iron up to the melting temperature, Sci. Rep. 4 (2014)
  5585.
\newblock \href {http://dx.doi.org/10.1038/srep05585}
  {\path{doi:10.1038/srep05585}}.

\bibitem{Carpenter1921}
H.~C.~H. Carpenter, C.~F. Elam, The production of single crystals of aluminium
  and their tensile properties, Proc. R. Soc. Lon. Ser. A 100~(704) (1921)
  329--353.
\newblock \href
  {http://arxiv.org/abs/http://rspa.royalsocietypublishing.org/content/100/704/329.full.pdf}
  {\path{arXiv:http://rspa.royalsocietypublishing.org/content/100/704/329.full.pdf}},
  \href {http://dx.doi.org/10.1098/rspa.1921.0089}
  {\path{doi:10.1098/rspa.1921.0089}}.

\bibitem{Kadeckova1967}
S.~Kadeckov\'{a}, B.~$\rm \check{S}$est\'{a}k,
  \href{http://dx.doi.org/10.1002/crat.19670020205}{{Growth of High-Purity Iron
  Single Crystals}}, Kris. Tech. 2~(2) (1967) 191--203.
\newblock \href {http://dx.doi.org/10.1002/crat.19670020205}
  {\path{doi:10.1002/crat.19670020205}}.
\newline\urlprefix\url{http://dx.doi.org/10.1002/crat.19670020205}

\bibitem{Lubitz1979}
K.~Lubitz, G.~G{\"o}ltz, \href{https://doi.org/10.1007/BF00932404}{{The
  preparation of large spherical iron single crystals}}, Appl. Phys. 19~(2)
  (1979) 237--239.
\newblock \href {http://dx.doi.org/10.1007/BF00932404}
  {\path{doi:10.1007/BF00932404}}.
\newline\urlprefix\url{https://doi.org/10.1007/BF00932404}

\bibitem{Behr2007}
G. Behr, W. L\"oser, Crystal Growth of Magnetic Materials (2007) In: Handbook
  of Magnetism and Advanced Magnetic Materials, John Wiley \& Sons, Ltd.

\bibitem{Gardner1978}
R.~Gardner, {Controlled growth of $\alpha$-Fe single crystal whiskers}, J.
  Cryst. Growth 43~(4) (1978) 425--432.
\newblock \href
  {http://dx.doi.org/http://dx.doi.org/10.1016/0022-0248(78)90340-8}
  {\path{doi:http://dx.doi.org/10.1016/0022-0248(78)90340-8}}.

\bibitem{Prinz1986}
G.~A. Prinz, B.~T. Jonker, J.~J. Krebs, J.~M. Ferrari, F.~Kovanic, {Growth of
  single crystal bcc $\alpha$-Fe on ZnSe via molecular beam epitaxy}, Appl.
  Phys. Lett. 48~(25) (1986) 1756--1758.
\newblock \href {http://arxiv.org/abs/http://dx.doi.org/10.1063/1.96778}
  {\path{arXiv:http://dx.doi.org/10.1063/1.96778}}, \href
  {http://dx.doi.org/10.1063/1.96778} {\path{doi:10.1063/1.96778}}.

\bibitem{Ardabili2005}
L.~Mohaddes-Ardabili, H.~Zheng, Q.~Zhan, S.~Y. Yang, R.~Ramesh,
  L.~Salamanca-Riba, M.~Wuttig, S.~B. Ogale, X.~Pan, {Size and shape evolution
  of embedded single-crystal $\alpha$-Fe nanowires}, Appl. Phys. Lett. 87~(20)
  (2005) 203110.
\newblock \href {http://arxiv.org/abs/http://dx.doi.org/10.1063/1.2128480}
  {\path{arXiv:http://dx.doi.org/10.1063/1.2128480}}, \href
  {http://dx.doi.org/10.1063/1.2128480} {\path{doi:10.1063/1.2128480}}.

\bibitem{Lu1997}
D.~L. Lu, K.~Tanaka, {Crystal habit of Ir, Rh, Co and Fe particles formed in
  solution}, J. Cryst. Growth 181~(4) (1997) 395--402.
\newblock \href
  {http://dx.doi.org/http://dx.doi.org/10.1016/S0022-0248(97)00304-7}
  {\path{doi:http://dx.doi.org/10.1016/S0022-0248(97)00304-7}}.

\bibitem{Zhang2006}
W.~Zhang, M.~Shimojo, M.~Takeguchi, K.~Furuya,
  \href{https://doi.org/10.1007/s10853-006-7783-1}{{Electron beam-induced
  formation of nanosized $\alpha$-Fe crystals}}, J. Mater. Sci. 41~(9) (2006)
  2577--2580.
\newblock \href {http://dx.doi.org/10.1007/s10853-006-7783-1}
  {\path{doi:10.1007/s10853-006-7783-1}}.
\newline\urlprefix\url{https://doi.org/10.1007/s10853-006-7783-1}

\bibitem{Uyeda1974}
R.~Uyeda, {The morphology of fine metal crystallites}, J. Cryst. Growth 24
  (1974) 69--75.
\newblock \href
  {http://dx.doi.org/http://dx.doi.org/10.1016/0022-0248(74)90282-6}
  {\path{doi:http://dx.doi.org/10.1016/0022-0248(74)90282-6}}.

\bibitem{Hayashi1977}
T.~Hayashi, T.~Ohno, S.~Yatsuya, R.~Uyeda,
  \href{http://stacks.iop.org/1347-4065/16/i=5/a=705}{{Formation of Ultrafine
  Metal Particles by Gas-Evaporation Technique. IV. Crystal Habits of Iron and
  Fcc Metals, Al, Co, Ni, Cu, Pd, Ag, In, Au and Pb}}, Jpn. J. Appl. Phys.
  16~(5) (1977) 705.
\newline\urlprefix\url{http://stacks.iop.org/1347-4065/16/i=5/a=705}

\bibitem{Canfield2001}
P.~C. Canfield, I.~R. Fisher, High-temperature solution growth of intermetallic
  single crystals and quasicrystals, J. Cryst. Growth 225~(2-4) (2001)
  155--161.
\newblock \href {http://dx.doi.org/10.1016/S0022-0248(01)00827-2}
  {\path{doi:10.1016/S0022-0248(01)00827-2}}.

\bibitem{Jesche2014c}
A.~Jesche, P.~C. Canfield,
  \href{http://dx.doi.org/10.1080/14786435.2014.913114}{Single crystal growth
  from light, volatile and reactive materials using lithium and calcium flux},
  Philos. Mag. 94~(21) (2014) 2372--2402.
\newblock \href {http://dx.doi.org/10.1080/14786435.2014.913114}
  {\path{doi:10.1080/14786435.2014.913114}}.
\newline\urlprefix\url{http://dx.doi.org/10.1080/14786435.2014.913114}

\bibitem{Jesche2014b}
A.~Jesche, R.~W. McCallum, S.~Thimmaiah, J.~L. Jacobs, V.~Taufour, A.~Kreyssig,
  R.~S. Houk, S.~L. Bud'ko, P.~C. Canfield,
  \href{http://dx.doi.org/10.1038/ncomms4333}{{Giant magnetic anisotropy and
  tunnelling of the magnetization in Li$_2$(Li$_{1-x}$Fe$_x$)N}}, Nat. Commun.
  5:3333, doi: 10.1038/ncomms4333.
\newline\urlprefix\url{http://dx.doi.org/10.1038/ncomms4333}

\bibitem{Westgren1922}
A.~Westgren, G.~Phragm\'en, X-ray studies on the crystal structure of steel, J.
  Iron Steel Inst. 105 (1922) 241--273.

\bibitem{jesche2016}
A.~Jesche, M.~Fix, A.~Kreyssig, W.~R. Meier, P.~C. Canfield,
  \href{http://dx.doi.org/10.1080/14786435.2016.1192725}{X-ray diffraction on
  large single crystals using a powder diffractometer}, Philos. Mag. 96~(20)
  (2016) 2115--2124.
\newblock \href {http://dx.doi.org/10.1080/14786435.2016.1192725}
  {\path{doi:10.1080/14786435.2016.1192725}}.
\newline\urlprefix\url{http://dx.doi.org/10.1080/14786435.2016.1192725}

\bibitem{Fuchs1965}
A.~Fuchs, B.~Ilschner,
  \href{https://doi.org/10.1107/S0365110X65003742}{{{Gitterkonstanten von
  Fe-Mo-Legierungen mit Mo-Gehalten bis 20 Gew.Prozent}}}, Acta Crystallogr.
  19~(3) (1965) 488.
\newblock \href {http://dx.doi.org/10.1107/S0365110X65003742}
  {\path{doi:10.1107/S0365110X65003742}}.
\newline\urlprefix\url{https://doi.org/10.1107/S0365110X65003742}

\bibitem{Virolainen2016}
S.~Virolainen, M.~F. Fini, V.~Miettinen, A.~Laitinen, M.~Haapalainen,
  T.~Sainio, {Removal of calcium and magnesium from lithium brine concentrate
  via continuous counter-current solvent extraction}, Hydrometallurgy 162
  (2016) 9--15.
\newblock \href
  {http://dx.doi.org/http://dx.doi.org/10.1016/j.hydromet.2016.02.010}
  {\path{doi:http://dx.doi.org/10.1016/j.hydromet.2016.02.010}}.

\bibitem{Wriedt1987}
H.~A. Wriedt, N.~A. Gokcen, R.~H. Nafziger,
  \href{https://doi.org/10.1007/BF02869273}{{The Fe-N (Iron-Nitrogen) system}},
  Bull. Alloy Phase Diagr. 8~(4) (1987) 355--377.
\newblock \href {http://dx.doi.org/10.1007/BF02869273}
  {\path{doi:10.1007/BF02869273}}.
\newline\urlprefix\url{https://doi.org/10.1007/BF02869273}

\bibitem{Wang2003}
W.~J. Wang, W.~X. Yuan, Y.~T. Song, X.~L. Chen, Thermodynamic calculation of
  the lithium-nitrogen system, J. Alloys Compd. 352~(1) (2003) 103--105.
\newblock \href
  {http://dx.doi.org/https://doi.org/10.1016/S0925-8388(02)01136-2}
  {\path{doi:https://doi.org/10.1016/S0925-8388(02)01136-2}}.

\bibitem{Crangle1971}
J.~Crangle, G.~M. Goodman, \href{http://www.jstor.org/stable/77809}{{The
  Magnetization of Pure Iron and Nickel}}, Proc. R. Soc. Lon. Ser. A 321~(1547)
  (1971) 477--491.
\newline\urlprefix\url{http://www.jstor.org/stable/77809}

\bibitem{Nakajima1990}
K.~Nakajima, S.~Okamoto, Large magnetization induced in single crystalline iron
  films by high‐dose nitrogen implantation, Appl. Phys. Lett. 56~(1) (1990)
  92--94.
\newblock \href {http://arxiv.org/abs/http://dx.doi.org/10.1063/1.102614}
  {\path{arXiv:http://dx.doi.org/10.1063/1.102614}}, \href
  {http://dx.doi.org/10.1063/1.102614} {\path{doi:10.1063/1.102614}}.

\bibitem{Coey1999}
J.~Coey, P.~Smith, Magnetic nitrides, J. Magn. Magn. Mater. 200~(1) (1999) 405
  -- 424.
\newblock \href
  {http://dx.doi.org/https://doi.org/10.1016/S0304-8853(99)00429-1}
  {\path{doi:https://doi.org/10.1016/S0304-8853(99)00429-1}}.

\bibitem{Sunagawa2010}
I. Sunagawa, Single Crystals Grown Under Unconstrained Conditions (2010) In:
  Dhanaraj G., Byrappa K., Prasad V., Dudley M. (eds) Springer Handbook of
  Crystal Growth. Springer, Berlin, Heidelberg.

\bibitem{Nolze2007}
G.~Nolze, {Image distortions in SEM and their influences on EBSD measurements},
  Ultramicroscopy 107~(2) (2007) 172--183.
\newblock \href
  {http://dx.doi.org/https://doi.org/10.1016/j.ultramic.2006.07.003}
  {\path{doi:https://doi.org/10.1016/j.ultramic.2006.07.003}}.

\bibitem{Massalski1996}
T.~B. Massalski (Ed.), Binary Alloy Phase Diagrams, Vol.~3, ASM International,
  (1996).

\bibitem{Lyublinski1995}
I.~E. Lyublinski, V.~A. Evtikhin, V.~Y. Pankratov, V.~P. Krasin, {Numerical and
  experimental determination of metallic solubilities in liquid lithium,
  lithium-containing nonmetallic impurities, lead and lead-lithium eutectic},
  J. Nucl. Mater. 224~(3) (1995) 288--292.
\newblock \href {http://dx.doi.org/10.1016/0022-3115(95)00076-3}
  {\path{doi:10.1016/0022-3115(95)00076-3}}.

\bibitem{Massalski1990b}
T.~B. Massalski (Ed.), Binary Alloy Phase Diagrams, Vol.~2, William W. Scott,
  (1990).

\end{thebibliography}

\end{document}